\documentclass[a4paper,twoside,10pt]{article}

\usepackage{coresa2024}

\usepackage[utf8]{inputenc}

\usepackage[T1]{fontenc}

\usepackage{xcolor}
\usepackage{amsfonts}
\usepackage{bm}
\usepackage{multirow}
\usepackage{booktabs}

\newcommand{\img}{\mathbf{x}}
\newcommand{\sysout}{\hat{\img}}
\newcommand{\latent}{\mathbf{y}}
\newcommand{\qlatent}{\hat{\latent}}
\newcommand{\synthparam}{\bm{\theta}}
\newcommand{\armparam}{\bm{\psi}}
\newcommand{\upparam}{\bm{\upsilon}}
\newcommand{\synth}{f_{\synthparam}}
\newcommand{\arm}{f_{\armparam}}
\newcommand{\upsampling}{f_{\upparam}}
\newcommand{\analysisparam}{\bm{\alpha}}
\newcommand{\analysis}{f_{\analysisparam}}

\definecolor{redsynthesis}{HTML}{E76F51}
\definecolor{greenarm}{HTML}{2CA395}
\definecolor{purpleupsampling}{HTML}{6600CC}
\definecolor{bluelatent}{HTML}{829CBC}
\definecolor{orangenonoverfitted}{HTML}{FF9933}

\graphicspath{{fig/}}

\begin{document}

\def\figurename{Figure}
\def\tablename{Table} 

\date{}

\title{\Large\bf N-O Cool-chic: reconcile fast encoding with lightweight decoding for neural image compression}

\author{Théophile Blard \\
      \textit{Orange Innovation}, France \\
      \texttt{\small{theophile.blard@orange.com}}}

\author{\begin{tabular}[t]{c@{\extracolsep{5em}}c}
            Théophile Blard, Théo Ladune, Pierrick Philippe & Xiaoran Jiang, Olivier Déforges                    \\
            \textit{Orange Innovation}, France              & \textit{IETR}, France                              \\
            \texttt{\small{firstname.lastname@orange.com}}  & \texttt{\small{firstname.lastname@insa-rennes.fr}}
      \end{tabular}}


\maketitle

\thispagestyle{empty}

\subsection*{Abstract}
{\em

      Overfitted image codecs achieve strong compression performance and low decoder complexity by learning a lightweight decoder for each image.
      Such codecs include Cool-chic, which presents image coding performance on par with VVC while requiring around 2000 multiplications per decoded pixel.
      However, the encoding time associated with overfitted codecs may be prohibitively long for real-time applications, posing a challenge to their practical implementation in such scenarios.
      To address this issue, this paper proposes to decrease the encoding complexity of Cool-chic by bypassing the overfitting procedure and complementing the decoder with an encoder network.
      The proposed non-overfitted (N-O) Cool-chic, significantly reduces encoding complexity by a factor of 1000 compared to Cool-chic, while maintaining competitive performance.
}
\subsection*{Index terms}
Neural image compression, low-complexity, overfitting

\section{Introduction}

Autoencoder-based codecs (ELIC \cite{elic}, MLIC++ \cite{mlicpp}) offer
state-of-the-art compression results, outperforming conventional codecs
(H.265/HEVC \cite{hevc}, H.266/VVC \cite{vvc}). During training, autoencoder
parameters are optimized following the rate-distortion cost computed on a
large dataset of images. Once the training stage
is completed, parameters are frozen and the autoencoder relies on
\textit{generalization} to compress unseen images. Generalization requires
networks with many parameters, making the decoding particularly complex. 
Indeed, autoencoder-based codecs have millions of parameters and require up to a million multiplications to decode a single pixel.
This decoding complexity might hinder their adoption especially when decoding happens on low-power devices such as smartphones.
\newline

Several studies have aimed to address the complexity constraint associated with autoencoder-based codecs.
For instance, Johnston et al. \cite{johnston2019computationally} achieved a 50\% reduction in the complexity of Ballé \cite{balle} through weight pruning applied to the decoder, without significant performance degradation.
EVC \cite{wang2023evc} leveraged network distillation to reduce the complexity by a factor of 10 while maintaining performance comparable to VVC.
More recently, Yang et al. \cite{DBLP:conf/iccv/YangM23} proposed the use of asymmetric architectures with shallow decoders, involving only 20~000 multiplications per decoded pixel, while still maintaining performance close to HEVC.
However, it is important to note that these approaches still exhibit significantly higher complexity compared to conventional methods.
\newline

\begin{figure}[t]
      \centering
      \includegraphics[width=\linewidth]{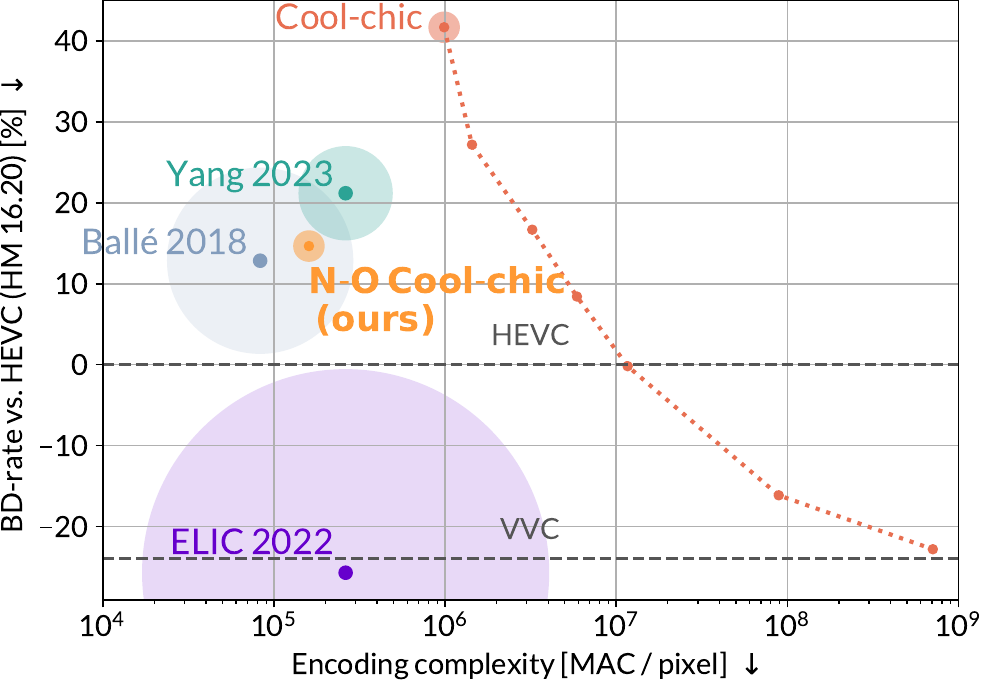}
      \caption{Rate-distorsion performance as a function of the encoding complexity on CLIC 2020 validation set \cite{clic20pro}.
            Negative results: less rate is required to get the same quality than HEVC.
            Cool-chic encoding complexity is varied by adjusting the training time.
            The circle radius denotes the decoding complexity (see Table \ref{tab:complexity}).}
      \label{fig:encoder-complexity-bdrate-clic}
\end{figure}

\begin{figure*}[t]
      \centering
      \includegraphics[width=\linewidth]{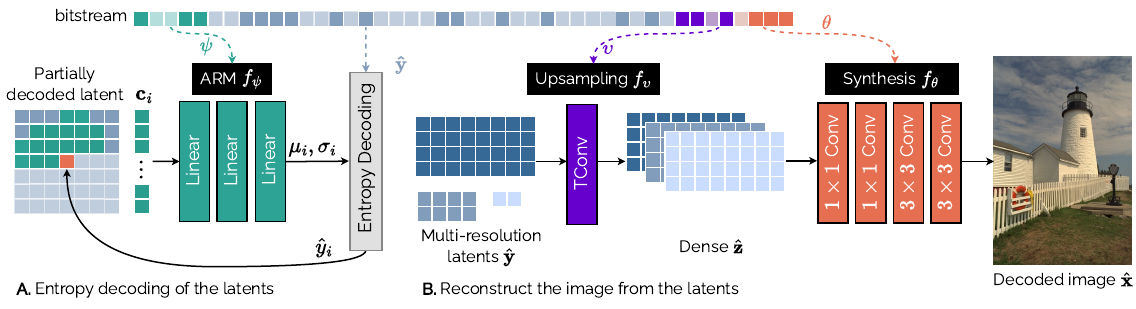}
      \caption{Cool-chic decoding. ARM: Auto-Regressive Model.}
      \label{fig:cool-chic-decoder}
\end{figure*}

Overfitted codecs (Cool-chic \cite{coolchic1, coolchic2, coolchic3}, C3 \cite{c3}) have
emerged as an alternative paradigm to autoencoders. To compress an image,
overfitted codecs learn (overfit) a lightweight neural decoder and latent
representation tailored for this image. The decoder parameters are then conveyed
alongside the latents so that the receiver can reconstruct the image.
Since overfitted codecs do not rely on generalization, their decoder are significantly lighter than autoencoders.
As such, they offer compelling image
coding performance on par with VVC with a decoder complexity of 2300
multiplications per pixel \cite{coolchic3}. 
As for conventional
codecs, this is obtained through an expensive encoding process, during which the
decoder and latents are optimized according to the image rate-distortion cost.

This paper aims to bridge the gap between autoencoder-based codecs and overfitted codecs
by combining the existing Cool-chic lightweight decoder with an encoder which generates the latent representation in a single forward pass.
The proposed network forms an autoencoder that offers the benefits of both low encoding and decoding complexity.

\section{Background: Cool-chic}
\label{sec:background}

This section presents the Cool-chic encoding and decoding process.
\newline

\textbf{Decoding.} Figure \ref{fig:cool-chic-decoder} presents the decoding
process of Cool-chic. It is composed of three main elements. \textit{i)} $L$ latent grids $\qlatent_l$ with different resolutions:
$\qlatent = \{
      \qlatent_{l} \in \mathbb{Z}^{H / 2^l \times W / 2^l},\ l = 0, \ldots, L - 1\}$.
\textit{ii)} The latent grids are transmitted using an entropy coding algorithm,
driven by an auto-regressive probability model $p_{\armparam}$ (ARM). This ARM
models the distribution of one latent value conditioned on neighbouring
values and is implemented as a MLP $\arm$. \textit{iii)} The upsampling
$\upsampling$ and synthesis $\synth$ networks upsample the latents to a
dense representation and synthesize the decoded image $\sysout$.
\newline

\textbf{Encoding.} Cool-chic encodes an image $\img$ by simultaneously learning
the latent and the different neural networks according to the image
rate-distortion cost:
\begin{align}
      \qlatent^*, \armparam^*, \upparam^*, \synthparam^*
       & = \operatorname*{argmin}_{\qlatent, \armparam, \upparam, \synthparam} \mathrm{D}(\img, \sysout) + \lambda \mathrm{R}(\qlatent) \label{eq:coolchic-encoding}                        \\
       & = \operatorname*{argmin}_{\qlatent, \armparam, \upparam, \synthparam} || \img - \synth(\upsampling(\qlatent)) ||^2 - \lambda \log_2 p_{\armparam} \left(\qlatent\right). \nonumber
\end{align}
The Lagrange multiplier $\lambda \in \mathbb{R}$ balances the rate $\mathrm{R}$
and the distortion $\mathrm{D}$, here the mean-squared error. The discrete
latents $\qlatent = \mathrm{Q}(\latent)$ are optimized through the continuous
version $\latent$, using the method proposed in C3 \cite{c3} to obtain a
differentiable proxy for the quantization $\mathrm{Q}$. After the encoding,
neural network parameters are quantized and entropy coded with an Exp-Golomb code 
since they usually represent less than 5~\%
of the total rate. The latents grids $\qlatent$ are entropy coded with a
range coder driven by the probability model $p_{\armparam}$.
\newline

\section{Non-overfitted Cool-chic}

Cool-chic encoding complexity can be mitigated at the expense of the compression performance by reducing the training (i.e. encoding) time.
While convenient, this is not enough for use-cases where \textit{real-time} encoding is required.
This section proposes a solution for situations where the encoding complexity constraint is paramount: a non-overfitted (N-O) Cool-chic.
\newline

\begin{figure*}[h]
      \centering
      \includegraphics[width=\linewidth]{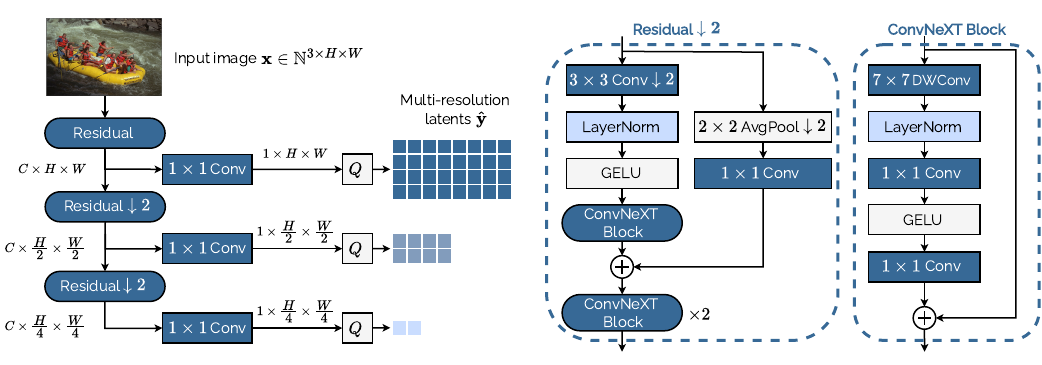}
      \caption{Proposed analysis transform. \textit{DWConv} stands for depth-wise convolution and $\downarrow$ denotes the stride.}
      \label{fig:cool-chic-analysis}
\end{figure*}

\textbf{Analysis transform. }In order to reduce the encoding time, the
overfitting process is bypassed by training a non-overfitted
(N-O) Cool-chic, sharing identical parameters for all images.
Taking inspiration
from autoencoders, encoding an image with N-O Cool-chic relies on an additional
analysis transform $\analysis$ generating a Cool-chic-compatible latent
representation $\qlatent$ from the input image $\img$:
\begin{equation}
      \qlatent = \mathrm{Q}(\analysis(\img)).
\end{equation}
As a result, the iterative optimization of $\qlatent$ with gradient descent is replaced with a single forward pass in the analysis network.
\newline

\textbf{Architecture overview. }The proposed analysis transform is depicted in Fig \ref{fig:cool-chic-analysis}.
To produce a set of $L = 7$ hierarchical latent
grids, a series of $L$ residual blocks progressively downsample the input image
$\img$ and extract relevant features for the different resolutions. After each
downsampling step, a $1\times 1$ convolution merges the $C = 64$ features
into the $l$-th latent $\qlatent_l$. The proposed analysis hence produces a few latent
grids with hierarchical resolutions, unlike autoencoder analysis which computes
hundreds of small-resolution features.
\newline

\textbf{Residual blocks. }To increase the receptive field while maintaining low
complexity, ConvNeXt blocks \cite{liu2022convnet} are adopted. 
Downsampling with residual block is achieved
by complementing the identity branch with a stride-2 $2\times 2$ average pooling
and a $1\times 1$ convolution as in \cite{he2019bag}. The first residual block does
not downsample so the pooling layer is removed. Following ELIC \cite{elic},
additional residual blocks are stacked after each downsampling for more
expressivity.
\newline

\textbf{Decoder. } The decoder parameters $(\armparam, \upparam, \synthparam)$ are learned alongside the analysis transform $\analysis$.
The architecture from \cite{coolchic3} with a complexity of 2300 MAC (multiplication-accumulation) per decoded pixel is selected.
It is worth noting that the network parameters no longer need to be transmitted alongside the latent representation since they are shared for all images.
\newline

\section{Experiments}

\textbf{Training. }The proposed N-O Cool-chic aims to obtain optimal parameters, generalizable to all possible images.
To this end, it is trained with $256\times256$ patches of randomly cropped images from the CLIC 2019 training set \cite{clic20pro}.
Since the global rate-distorsion cost must be optimized, equation \eqref{eq:coolchic-encoding} becomes:
\begin{align}
      \analysisparam^*, & \armparam^*, \upparam^*, \synthparam^*
      = \operatorname*{argmin}_{\analysisparam, \armparam, \upparam, \synthparam} \mathbb{E}_{\img} [\mathrm{D}(\img, \sysout) + \lambda \mathrm{R}(\qlatent)] \label{eq:non-overfitted-coolchic-encoding}                              \\
                        & = \operatorname*{argmin}_{\analysisparam, \armparam, \upparam, \synthparam} \mathbb{E}_{\img} [ || \img - \synth(\upsampling(\qlatent)) ||^2 - \lambda \log_2 p_{\armparam} \left(\qlatent\right)]. \nonumber
\end{align}
Independent training with different rate constraints is performed to cover a
wide range of rates, namely $\lambda = \{ 0.02, 0.004, 0.001, 0.0004, 0.0001
      \}$. Adam algorithm \cite{kingma2014adam} is used, the learning rate starts from
$10^{-3}$ and is dynamically decayed by a patience mechanism. When it reaches
$10^{-6}$, the training is terminated.
\newline

\textbf{Encoding complexity. }Figure \ref{fig:encoder-complexity-bdrate-clic} shows the
compression performance of N-O Cool-chic against the encoding complexity.
N-O Cool-chic encodes images in a single forward pass through the analysis.
This reduces the complexity by a factor of 1000 compared to Cool-chic, requiring only 160 kMAC
per pixel which is comparable to other autoencoder-based encoders.
While reducing the training time can help decrease the encoding complexity of Cool-chic, it still requires 20 times more MACs compared to N-O Cool-chic to reach the same level of performance.
Beside the reduction in MAC, encoding images with N-O Cool-chic is conceptually simpler
since there is no more optimization through gradient descent. However, this
reduces the performance significantly, falling behind state-of-the art codecs such as ELIC \cite{elic}.
\newline

\begin{figure}[t]
      \centering
      \includegraphics[width=\linewidth]{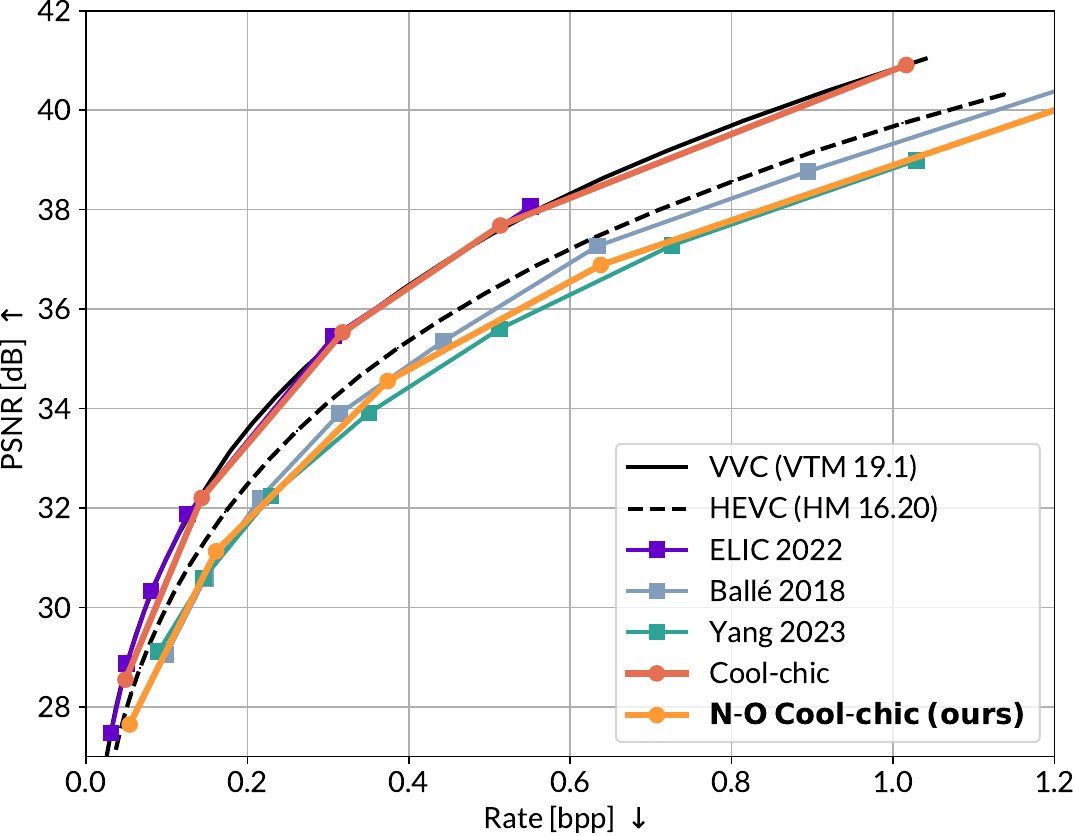}
      \caption{Rate-distortion performance on the CLIC 2020 validation set \cite{clic20pro}.}
      \label{fig:rd-curve-clic20}
\end{figure}

\begin{table*}[ht]
      \centering

      \begin{tabular}{cccc}
            \hline
            \multirow{2}{*}{Method}                 & \multicolumn{2}{c}{Complexity (kMAC / pixel)} & \multirow{2}{*}{BD-Rate vs HEVC (\%) $\downarrow$}         \\
            \cline{2-3}
                                                    & Encoder                                       & Decoder                                                    \\
            \hline
            ELIC 2022 \cite{elic}                   & 262.2                                         & 382.0                                              & -25.7 \\
            Ballé 2018 \cite{balle}                 & 80.0                                          & 83.0                                               & 12.9  \\
            Yang 2023 \cite{DBLP:conf/iccv/YangM23} & 262.2                                         & 20.5                                               & 21.2  \\
            N-O Cool-chic (ours)                    & 160.0                                         & \textbf{2.3}                                       & 14.7  \\
            \hline
      \end{tabular}
      \caption{Encoder and decoder complexity \textit{v.s.} average BD-rate relative to HEVC on CLIC 2020 validation set \cite{clic20pro}.}
      \label{tab:complexity}
\end{table*}

\textbf{Rate-distorsion results. } Figure \ref{fig:rd-curve-clic20} presents the rate-distortion curves.
Compared to Cool-chic, N-O Cool-chic requires a 45\% higher rate to achieve the same quality.
Compared to Ballé \cite{balle}, this gap is reduced to only 1\%, with similar encoding complexity but with a decoder that is 30 times less complex (Table \ref{tab:complexity}).
On the other hand, N-O Cool-chic surpasses models that are specifically optimized for low complexity decoding, such as the 2-layer synthesis model proposed in \cite{DBLP:conf/iccv/YangM23}.
It achieves the same quality with a 6\% lower rate, while also having a decoder that is 8 times less complex.

\section{Conclusion}

This paper strives to make Cool-chic encoding faster. 
To this end N-O Cool-chic is proposed, where the encoding becomes a simple forward pass (less than 1 second), reducing the encoding complexity up to a factor of 1000. 
It is shown that although the decoding complexity is only 2300 multiplications per pixel, N-O Cool-chic achieves comparable performance to recent neural image codecs.
However, it falls short of the performance achieved by the overfitted Cool-chic, underscoring the crucial role of overfitting in attaining optimal performance and adaptation.



\bibliographystyle{unsrt}
\bibliography{coresa2024}

\end{document}